%% file: paper.tex
\documentclass[11pt]{article}

\input{arxiv_style}
\input{dylan}

\usepackage{float}
\usepackage{subcaption}
\usepackage{amsmath}
\newcommand{\defeq}{\stackrel{\text{def}}{=}}

\title{\textbf{Universal Supervised Learning for Individual Data}}
\date{}

\author{Yaniv Fogel and Meir Feder
\thanks{School of Electrical Engineering,
	Tel Aviv University, Tel Aviv, 6887801, Israel}
}

\begin{document}

\maketitle

\begin{abstract}
  \input{abstract}

\end{abstract}

\section{Introduction}
\label{sec:introduction}
\input{section_introduction}

\section{Problem Definition}
\label{sec:problem_statement}
\input{section_problem_statement}

\section{Main Results}
\label{sec:main_results}
\input{section_main_results}

\section{The pNML: Additional Interpretation and Extensions}
\label{sec:pnml_interpretation}
\input{section_pnml_interpretation}


\section{Simple Examples}
\label{sec:simulation_results}
\input{section_simulation_results}

\section{Conclusions and Discussion}
\label{sec:conclusions}
\input{conclusions}

\section*{Acknowledgments}
Koby Bibas is acknowledged for discussions and for implementing and analyzing the pNML in various problems, from linear regression to deep neural networks. The joint work with Koby appears in \cite{pNML_linear_regression,pNML_neural_networks}.

We also acknowledge the discussion with Amichai Painsky regarding section \ref{sec:pnml_interpretation}, and the related work \cite{PainskyFeder1}.

{
\bibliography{references}
}

\appendix
\section{The Stochastic Setting}
\input{AppendixA}
\section{Twice Universality}
\input{AppendixB}

\end{document}

%% file: arxiv_style.tex
\usepackage[letterpaper, left=1in, right=1in, top=1in, bottom=1in]{geometry}

\usepackage{parskip}

\usepackage[dvipsnames]{xcolor}
\usepackage[colorlinks=true, linkcolor=blue, citecolor=blue, pagebackref, bookmarks=true]{hyperref}
\usepackage{bookmark}
\usepackage{microtype}

\usepackage{algorithm}
\usepackage{algpseudocode}

\usepackage{natbib}
\bibpunct{(}{)}{;}{a}{,}{,}

\usepackage{amsthm}
\usepackage{mathtools}
\usepackage{amsmath}
\usepackage{bbm}
\usepackage{amsfonts}
\usepackage{amssymb}

\usepackage{MnSymbol} 

\theoremstyle{definition}  

\theoremstyle{plain}

\usepackage{prettyref}

\newcommand{\savehyperref}[2]{\texorpdfstring{\hyperref[#1]{#2}}{#2}}
\newrefformat{eq}{\savehyperref{#1}{\textup{(\ref*{#1})}}}
\newrefformat{eqn}{\savehyperref{#1}{Equation~\ref*{#1}}}
\newrefformat{lem}{\savehyperref{#1}{Lemma~\ref*{#1}}}
\newrefformat{def}{\savehyperref{#1}{Definition~\ref*{#1}}}
\newrefformat{line}{\savehyperref{#1}{line~\ref*{#1}}}
\newrefformat{thm}{\savehyperref{#1}{Theorem~\ref*{#1}}}
\newrefformat{corr}{\savehyperref{#1}{Corollary~\ref*{#1}}}
\newrefformat{sec}{\savehyperref{#1}{Section~\ref*{#1}}}
\newrefformat{app}{\savehyperref{#1}{Appendix~\ref*{#1}}}
\newrefformat{ass}{\savehyperref{#1}{Assumption~\ref*{#1}}}
\newrefformat{ex}{\savehyperref{#1}{Example~\ref*{#1}}}
\newrefformat{fig}{\savehyperref{#1}{Figure~\ref*{#1}}}
\newrefformat{alg}{\savehyperref{#1}{Algorithm~\ref*{#1}}}
\newrefformat{rem}{\savehyperref{#1}{Remark~\ref*{#1}}}
\newrefformat{conj}{\savehyperref{#1}{Conjecture~\ref*{#1}}}
\newrefformat{prop}{\savehyperref{#1}{Proposition~\ref*{#1}}}
\newrefformat{proto}{\savehyperref{#1}{Protocol~\ref*{#1}}}
\newrefformat{prob}{\savehyperref{#1}{Problem~\ref*{#1}}}
\newrefformat{claim}{\savehyperref{#1}{Claim~\ref*{#1}}}

%% file: dylan.tex
\let\P\undefined

\DeclareMathOperator{\P}{P}

\DeclareMathOperator*{\argmax}{arg\,max}

\def\ddefloop#1{\ifx\ddefloop#1\else\ddef{#1}\expandafter\ddefloop\fi}
\def\ddef#1{\expandafter\def\csname bb#1\endcsname{\ensuremath{\mathbb{#1}}}}
\ddefloop ABCDEFGHIJKLMNOPQRSTUVWXYZ\ddefloop
\def\ddefloop#1{\ifx\ddefloop#1\else\ddef{#1}\expandafter\ddefloop\fi}
\def\ddef#1{\expandafter\def\csname b#1\endcsname{\ensuremath{\mathbf{#1}}}}
\ddefloop ABCDEFGHIJKLMNOPQRSTUVWXYZ\ddefloop
\def\ddef#1{\expandafter\def\csname c#1\endcsname{\ensuremath{\mathcal{#1}}}}
\ddefloop ABCDEFGHIJKLMNOPQRSTUVWXYZ\ddefloop
\def\ddef#1{\expandafter\def\csname h#1\endcsname{\ensuremath{\widehat{#1}}}}
\ddefloop ABCDEFGHIJKLMNOPQRSTUVWXYZ\ddefloop
\def\ddef#1{\expandafter\def\csname hc#1\endcsname{\ensuremath{\widehat{\mathcal{#1}}}}}
\ddefloop ABCDEFGHIJKLMNOPQRSTUVWXYZ\ddefloop
\def\ddef#1{\expandafter\def\csname t#1\endcsname{\ensuremath{\widetilde{#1}}}}
\ddefloop ABCDEFGHIJKLMNOPQRSTUVWXYZ\ddefloop
\def\ddef#1{\expandafter\def\csname tc#1\endcsname{\ensuremath{\widetilde{\mathcal{#1}}}}}
\ddefloop ABCDEFGHIJKLMNOPQRSTUVWXYZ\ddefloop


%% file: abstract.tex

 Universal supervised learning is considered from an information theoretic point of view following the universal prediction approach, see \cite{universal_prediction}. 
 We consider the standard supervised ``batch'' learning where prediction is done on a test sample once the entire training data is observed, and the individual setting where the features and labels, both in the training and test, are specific individual quantities. 
 The information theoretic approach naturally uses the self-information loss or log-loss. 

Our results provide universal learning schemes that compete with a ``genie'' (or reference) that knows the true test label. 
In particular, it is demonstrated that the main proposed scheme, termed Predictive Normalized Maximum Likelihood (pNML), is a robust learning solution that outperforms the current leading approach based on Empirical Risk Minimization (ERM). Furthermore, the pNML construction provides a pointwise indication for the learnability of the specific test challenge with the given training examples. 

%% file: section_introduction.tex

The common situation in supervised learning is as follows. A set of training examples is given, composed of, say, $N$ samples of pairs $\{z_i=(x_i,y_i)\}_{i=1}^{N}$, where $x_i\in {\cal X}$ denotes the $i$-th feature (usually ${\cal X}$ is a large space, e.g., the space of images, large vector of features and so on) and $y_i\in {\cal Y}$ is its corresponding label (${\cal Y}$ is usually a finite small set, but may take continuously many values in regression problems). 
Then, a test feature $x$ is shown, and the machine learning task is to predict the corresponding $y$. 
The prediction may be a value $b=\hat{y}=f(x)$, but may also be a weighting $b=q(\cdot|x)$ for all $y\in {\cal Y},\; q(y|x)\geq 0, \;\sum_y q(y|x)=1$, which can be regarded as a probability assignment. There is a loss function $\ell(b,y)$ in making the prediction. In our information theoretic analysis, we consider the log-loss, \[\ell(b,y|x) = -\log q(y|x).\] 
There are many reasons why to use the log-loss which will not be repeated here (see, e.g., \cite{painsky2018universality}), we will only mention the fact that by considering log-loss we can obtained an elegant, information theoretical framework for the learning/prediction problem.

Since the prediction is generated based on the training, we sometimes express this explicitly: $$q(\cdot|x;x^{N},y^{N})=q(\cdot|x;z^{N})$$ where we use the notation $x^L=x_1,\ldots,x_L$. 

Clearly, the ultimate goal is to minimize the loss. But this task is not clear if we do not make additional assumptions regarding the way the data is generated or on the class of possible ``models'' or ``hypotheses'' used in order to find the relation between $x$ and $y$.

\subsection*{The Model Class} 
The model class definition plays an important role 
in all settings we consider. 
Specifically, a model class is 
a set of conditional probability distributions
\[ P_\Theta = \{ p_\theta(y|x),\;\;\theta\in\Theta\} \] 
where $\Theta$ is a general index set. This is equivalent to saying that there is a set of stochastic functions  $\{ y=g_\theta(x),\;\;\theta\in\Theta\}$ used to explain the relation between $x$ and $y$.

A major issue is how to choose a model class. As common sense indicates, on one hand one may wish to choose a large as possible class, so that any possible relation between $x$ and $y$ can be captured by some member in the class. However, if the class is too large, it may not be ``learnable''. That is, it will be impossible to deduce reliably on the large class based on the finite training example of size $N$. This notion appears in classical statistical reasoning and expressed, e.g., as the bias-variance trade-off. This major issue of choosing the model class will be discussed briefly towards the end of the paper, but throughout the paper we assume that $P_\Theta$ is given. 

\subsection*{The Various Settings}
In the different settings used to come up with a solution to the above learning problem, we make different assumptions on how the data is generated. The simplest is the stochastic setting where it is assumed that the data (and the relation between $x$ and $y$) is indeed generated by a model $p_\theta(y|x)$  which is an unknown member of $P_\Theta$. The learning problem in the stochastic setting has been analyzed in \cite{FogelFederISIT2018}, and 
is briefly described in Appendix \ref{appendix a}.
In addition to the propsed learner, in that work we have identified the information theoretical ``capacity'', $C=\max_{w(\theta)} I(\Theta;Y|Y^N,x^N,x)$ where $I$ is the mutual information between the class $\Theta$, now a random variable with a distribution $w(\cdot)$ attaining the capacity, and the label $Y$, given the training and test feature, where the labels of the training are random variables as they are generated by some $p_\theta(y|x)$. In the stochastic setting $C$ defines if the class of models is learnable, if it vanishes with the training size $N$.

The next setting is the most commonly used in the learning community and is known as the Probably Approximately Correct (PAC) setting.
The concept of PAC-learning was established in the famous \cite{Valiant1984} work.
In PAC it is assumed that the data samples are generated independently from some source $P(x,y)=P(x)P(y|x)$. Unlike the stochastic setting, $P(y|x)$ is not necessarily a member of the hypotheses class.
The common purpose in PAC is to design algorithms that attain, with high probability over the possible training sets (hence 'probably'), a loss which is almost as that of the best hypothesis in the class (hence 'approximately'). 

Following this PAC setting, several measures of the class learnability have been suggested, such as the Vapnik-Chervonenkis (VC) dimension \cite{VC}, the Rademacher complexity \cite{bartlett2002rademacher} and more. These measures and the theory behind them successfully explain the learnability of rather simple hypotheses classes, yet they seem to fail in explaining the learnability of modern model classes, most importantly deep neural networks (DNN), see \cite{zhang2016understanding}. This gap between the theoretical learnability measures and empirical results need to be explored further. Some answer to this discrepancy is given by our work below.  

The strongest setting, however, of universal learning is the individual setting where there is no assumption on the way the data
is generated: the training $z^{N}$ is an individual, specific set,
and then a specific test point $x$ is presented to the learner. The
goal is to predict the corresponding label $y$ for the test
sample, which is again a deterministic unknown. As common in this setting, the learner $q(\cdot|x; z^{N})$
should be chosen so that it can reasonably compete with a ``genie'' that
knows the specific sequence, yet constrained to use an explanation
from the model class $P_\Theta$. This leads to a minmax approach where
the minimum is over the choice of $q$ and the maximum is with respect
to any possible outcome.

Interestingly, there are several ways to define the reasonable criterion which will be elaborated in the sequel. Our problem definition is the crucial aspect of the proposed solution. 
It turns out that our resulting schemes, that have precise optimality
criteria, are  more ``stable'' than the standard approach based on
empirical risk minimization (ERM) using the stochastic gradient
(SGD). The main solution we advocate is termed Predictive/Pointwise Normalized Maximum Likelihood (pNML). This solution was proposed before for the related universal prediction problem, see \cite{roos2008sequentially,roos2008bayesian}, as an efficient implementation of Shtarkov's Normalized Maximum Likelihood method \cite{shtar1987universal}. 
Furthermore, the pNML provides an information theoretic
learnability measures that depends on the specific training $z^N$
and the specific test features $x$ and 
may let the learner {\em know when it does not know.} 
The advantages of our schemes, especially the pNML and its learnability measure, are demonstrated in this paper in some simple examples and in follow-up works \cite{pNML_linear_regression,pNML_neural_networks} 
in a variety on learning situations from basic linear regression
to deep neural networks (DNN).

The proposed learnability measure resembles similar criteria suggested in the learning community based on the concept of stability. Specifically, contrary to Rademacher complexity and VC-dimension, which are measures of the richness of an hypotheses class, stability is a measure a learning algorithm. Broadly speaking, an algorithm is stable if the effect of changing one of the training examples can be bounded. The stability of a learning algorithm can be used to bound the generalization error, see \cite{bousquet2002stability} and \cite{shalev2010learnability}. However, while our information theoretical measure resembles stability measures, it is different, as it is not based on perturbing the training, but on checking all possible behaviors of the unknown test label. 

\subsection*{Universal Prediction and Online Learning}
Before moving on to the presentation of our results, we mention the problem of on-line learning, which is closer to the well studied work on universal prediction \cite{universal_prediction}. The latter is essentially an online learning but with no $x$'s. As described in \cite{universal_prediction}, the universal prediction solution with log-loss provides a ``universal probability'' for the entire sequence, i.e., $q(y^N)$, which can be converted to a sequential prediction strategy via the chain rule $q(y^N)=\prod_{t=1}^N q(y_t|y^{t-1})$. The universal probability is given by either a Bayesian mixture $q(y^N) = \int_\Theta w(\theta) p_\theta (y^N)$ (in the stochastic setting) or the normalized maximum likelihood (NML) $q_{\mbox{\tiny{NML}}}(y^N) = \frac{\max_{\theta} p_{\theta}(y^N)}{\sum_{\tilde{y}^N}\max_{\theta} p_{\theta}(\tilde{y}^N)}$ (in the individual setting). Both solutions solve a corresponding minmax problem. So universal prediction is well understood.

Online learning, with a feature sequence $x^N$ is less understood as it has one important difference: the chain rule does not exist for conditional probabilities, that is not every probability $p(y^N|x^N)$ can be factored in a product of conditional sequential probabilities $p(y_t|x_t; y^{t-1}, x^{t-1})$. Indeed, at least in the individual setting, one can come up with simple examples where the reference ``genie'' can assign a perfect match for a given individual sequence, yet sequentially any universal assignment will fail. The problem of on-line learning, especially in the individual setting has been recently analyzed in \cite{fogel2017problem}. But even after this work the full solution of the on-line learning problem is still open.

\subsection*{Paper Structure}
While online learning is still open, in this paper we focus on the standard ``batch'' supervised learning problem, in the individual setting. The paper is organized and presents the results as follows.
In section \ref{problem_statement} we present two possible definitions of the batch learning problem in the individual setting. The solution of the corresponding min-max problem for the first definition, denoted pNML, and an upper bound for the second definition, are given in section \ref{MainResult}. 
In section \ref{sec:pnml_interpretation}, we provide additional interpretation and extensions to the pNML.
Experimental results for simple learning problems are given in section \ref{SimulationResults}. Finally, section \ref{conclusions} concludes the paper, discusses alternative definitions, considers the model class selection problem and offers a twice universal solution (elaborated in the Appendix) and suggests future work.


%% file: section_problem_statement.tex
\label{problem_statement}
This paper takes the individual setting where it is assumed that both the training $z^N=(x^{N},y^{N})$ and test $z=(x,y)$ (both the fearure test vector and the corresponding label) are specific individual entities. We will sometimes denote $z_{N+1}=z,\;z^{N+1}=z^N,z$.

In the learning problem these specific training examples $z^{N}$ as well as the test feature vector $x$ are given to the learner before its decision on the prediction. So it seems that there is no harm, and even an advantage, in assuming that these data values are specific, deterministic, provided that a reasonable criterion can be suggested for the prediction accuracy of the unknown label. Such a criterion may depend on the given data beforehand, but should lead to a good desired performance no matter what the unknown label is. This is the essence of our problem formulation.

The learner in our setting predicts a distribution for the unknown label denoted $q(\cdot|x;z^{N})$.
Any universal learner is a pre-designed procedure for generating such distributions. So what can be a good procedure for the learner when the test outcome $y$ is arbitrary and may even be chosen by an adversary that knows the learner procedure? As it is common in the individual setting in the field of universal prediction, the learner can only compare itself to a ``genie'', or a reference learner that knows the label value, but is constrained to use an assignment from the given class of models $P_\Theta$. It turns out that this is not enough. In the batch, one-time test, such a ``genie'' is too powerful for any learner to compete against. Thus, we suggest that the reference learner would indeed be a procedure from $P_\Theta$, would know the entire sequence (including the test feature and label) $z^N,x,y$, but would not know which instance $(x_t,y_t)$ out of the $N+1$ pairs is the test. Thus, the reference learner chooses the member $\hat{\theta}=\hat{\theta}(x^N,y^N,x,y)$ that minimizes the average log-loss over the entire sequence that includes both the training and the test samples:
\begin{equation}
\label{reference}
\hat{\theta}(z^N,x,y) = \arg \max_{\theta} p_{\theta}(y^N,y|x^N,x) = \arg \max_{\theta} \left[ p_\theta(y|x)\prod_{t=1}^{N} p_{\theta}(y_t|x_t) \right]
\end{equation}
where the last equality holds under the common assumption that the model class distribution for a set of measurements is i.i.d.
With this reference, we can define the following regret:

\begin{align}
R_{ind}(q,x,y;z^{N}) = \log\left(\frac{p_{\hat{\theta}(z^N,x,y)}(y|x)}{q(y|x;z^{N})}\right)
\end{align}
Since $y$ is not known, we propose $q$ which minimizes the regret for any $y$, i.e., the minimax regret:

\begin{align}
\label{pnml_regret}
R^*_{ind}(x,z^{N}) = \min_{q} \max_{y} R_{ind}(q,x,y;z^{N}).
\end{align}

This seems like a natural definition for the learning problem, since in real life applications we are given a specific training set and the test feature. This perspective is different from the standard statistical learning theory settings, where one derives bounds that should apply to 'most' of the training sets. In addition, the resulting regret explicitly depends on the obsereved test feature, which may allow us to distinguish between learnable and unlearnable test features. This is also contrary to classical statistical learning theory where the performance measure is an average over the possible test features. 

Note that there may be issues with the definition of the benchmark loss, since in general there may be several hypotheses $\theta \in \Theta$ that achieve the minimal loss (\ref{reference}). Those hypotheses might assign different probabilities for the test outcome, and so the benchmark with which the learner has to compete is not clearly defined. This can be solved, for example, by taking the average over all $\theta$'s that achieve the maximum likelihood $\hat{\theta}$. Nevertheless, in most examples we may assume that there is a single minimizer for $-\log \left({\theta}(y^N,y|x^N,x) \right)$ for all possible $z^N,x,y$, thus avoiding this issue, or that we use some specific minimizer (a particular training procedure) that converges to a single specific $\hat{\theta}$.

Another, more serious issue with the criterion (\ref{pnml_regret}) is that it may be considered ``unfair'': While the reference does not know which of the points is the test point, the universal learner does. Also, the above criterion may seem too dependent on the particular $x$, and the solution may be biased accordingly. 
Thus, as an alternative, and to avoid such a dependency, one may consider a definition in which the regret, to be minmax-ed, is calculated over all permutations of the individual sequence of training and test:

\begin{align}
\label{loo_regret}
R_{perm}(q,z^{N+1}) = \frac{1}{(N+1)!}\sum_{\tilde{z}^{N+1} = perm(z^{N+1})}\log\left(\frac{p_{\hat{\theta}(z^{N+1})}(\tilde{y}_{N+1}|\tilde{x}_{N+1})}{q(\tilde{y}_{N+1}|\tilde{x}_{N+1};\tilde{z}^{N})}\right)
\end{align}
which, for the case where $p_{\theta}(y^{N+1}|x^{N+1})=\prod_{t=1}^{N+1} p_{\theta}(y_t|x_t)$, is as follows:
\begin{align}
R_{perm}(q,z^{N+1}) = R_{LOO}(q,z^{N+1}) = \frac{1}{N+1}\sum_{t=1}^{N+1} \log\left(\frac{p_{\hat{\theta}(z^{N+1})}(y_t|x_t)}{q(y_t|x_t;z^{(N+1)\setminus t})}\right)
\end{align}
where $(N+1)\setminus t$ means all the indices $1,\ldots,N+1$ besides $t$, and $LOO$ stands for ``Leave One Out", which is essentially what is suggested here: to take $N+1$ experiments over the sequence where each time one point is left out as a test, and consider the average, empirical regret over these experiments.  
The permutation or $LOO$ also solves the difficulty that arise when there are several hypotheses that achieves the minimal loss, since the benchmark against which the learner competes is the sum of the losses, $\log p_{\hat{\theta}(z^{N+1})}(y^{N+1}|x^{N+1})=\sum_{t=1}^{N+1} \log p_{\hat{\theta}(z^{N+1})}(y_t|x_t)$.

Again, we will be interested in the min-max regret, where the minimization is over any procedure for assigning $q$ that depends on the current feature $x_t$ and $N$ training examples $z^{(N+1)\setminus t}$, and the maximization should now be over all $y^{N+1}$:
\begin{align}
\label{perm_regret}
R^*_{perm}(x^{N+1}) = \min_q \max_{y^{N+1}} R_{perm}(q,z^{N+1})
\end{align}
In some sense, the resulting $q$ controls the worst case regret averaged over all data points.


%% file: section_main_results.tex
\label{MainResult}
\subsection*{Pointwise solution: Predictive Normalized Maximum Likelihood}
	
In the problem (\ref{pnml_regret}) the optimization is ``pointwise'': for each training $z^{N}$ and test feature value $x$.
It is rather straightforward to see, using the equalizer reasoning, that the resulting minimax optimal probability assignment $q$ is:
\begin{align}
	\label{individual_solution}
	q(y|x; z^{N}) = \frac{p_{\hat{\theta}(z^N,x,y)}(y|x)}{\sum_{y'}p_{\hat{\theta}(z^N,x,y'}(y'|x)} = q_{\mbox{\tiny{pNML}}}(y|x; z^{N}).
\end{align}
We call this solution the Predictive/Pointwise Normalized Maximum Likelihood (pNML) learner. 

We note that the pNML probability assignment (\ref{individual_solution}) was proposed earlier, see \cite{roos2008sequentially,roos2008bayesian}, as one of a possible variations of the known Normalized Maximum Likelihood (NML) method of \cite{shtar1987universal} for universal prediction. However, it was suggested with a different motivation, mainly to simplify the evaluation of the NML for universal prediction. This assignment and other variations of the NML were also mentioned in the book \cite{Grunwald2007} as sequential NML (SNML). The previous work also noted that this probability assignment is the solution to the minmax problem (\ref{pnml_regret}) (with no $x$'s for prediction), but  in our formulation the minmax problem did come with a motivation - to compete with the leave-one-out reference.

To prove that this is indeed the minimax solution, note that the regret is equal for all choices of $y$. Now, if we consider a different probability assignment, it should assign a smaller probability for at least one of the possible outcomes. In this case, choosing one of those outcomes will lead to a higher regret. 

The induced min-max regret is:
\begin{align}
\label{individual_regret}
R^*_{ind}(z^{N},x) = \log\left(\sum_{y}p_{\hat{\theta}(z^N,x,y)}(y|x) \right)\defeq \Gamma_{\mbox{\tiny{pNML}}}(z^N,x),
\end{align}
which, as expected, is greater then zero:
\begin{align}
\label{individual_regret_positive}
\log\left(\sum_{y}p_{\hat{\theta}(z^N,x,y)}(y|x)\right) \geq \log\left(\sum_{y}p_{\hat{\theta}(z^N,x,y^*)}(y|x)\right) = 0 
\end{align}
where we used the fact that the likelihood adjusted to each $y$ in the sum is greater than the likelihood associated with any fixed $y^*$ (which may even be the true label), i.e., $p_{\hat{\theta}(z^N,x,y^*)}(y|x) \leq p_{\hat{\theta}(z^N,x,y)}(y|x)$, and that the sum over all possible $y$'s of probability distribution on $y$ associated with a fixed $\theta = \hat{\theta}(z^N,x,y^*)$ is $1$.

If the choice of the next unknown label $y$ does not change the maximum likelihood (i.e., $\hat{\theta}$ is not affected by the choice of $y$), the regret is $0$. On the other hand, if the hypotheses class is such that the maximum-likelihood $\hat{\theta}$ varies considerably with choice of $y$, i.e., the sensitivity of $\hat{\theta}=\hat{\theta}(z^N,x,y)$ to the choice of a new label $y$ (while the rest $z^N$ and $x$ are fixed) is high, the regret will be large. Note that $\Gamma_{\mbox{\tiny{pNML}}}(z^N,x)$ indicates a local, pointwise behavior at $z^N,x$, via the behavior of the particular function $f_{(z^N,x)}(y) = \hat{\theta}(z^N,x,y)$, and not a global behavior of the model class $P_\Theta$. This regret, or log-normalization factor $\Gamma$, can thus serve as a learnability measure at the particular training and test example: when it is small, the resulting learner is reliable and the cost of universality is small, while when $\Gamma$ is large the uncertainty in the learning ishigh and thus a large cost of universality is required.

We note that this interpretation may resemble several stability notions which were studied in the past, see e.g., \cite{shalev2010learnability},\cite{bousquet2002stability}. Contrary to those notions, however, here we do not leave out or change one of the training examples, but instead add each of the possible new label values for the specific data feature on which we are tested.

\subsection*{Permutation and Leave One Out Average}

Moving on to the analysis of the regret in the permuted scenario, we look for the solution of (\ref{perm_regret}). Unfortunately, we were not able to solve this min-max problem. Nevertheless, we will present several upper bounds for $R^*_{perm}(x^N,x)$, that are achieved by considering various possible probability assignments $q$. 

To simplify notation, let us consider probability assignments $q$'s that are interchangeable with respect to the training set, i.e., for all $(x,y)^{N}$ and $(\tilde{x},\tilde{y})^{N}$ that are the same up to a permutation, the following holds: 

\begin{align}
q(y|x;x^{N},y^{N}) = q(y|x,\tilde{x}^{N},\tilde{y}^{N}).
\end{align} 

Narrowing down our potential probability assignments will allow us to upper bound the min-max regret. 
If in addition the models in the class $P_\Theta$ are also exchangeable (or i.i.d), the regret of the permutation is the average leave-one-out regret, given by (\ref{loo_regret}). 

Now, to get an upper bound for this regret, let us consider a specific, easy to analyze probability assignment, denoted $q_{\tiny{\mbox{NML}}}$, that coincides with the normalized maximum likelihood principle over the entire sequence (including the test):  

\begin{align}
\label{NML}
q_{\tiny{\mbox{NML}}}(y_t|x_t; z^{(N+1)/t}) = \frac{p_{\tiny{\mbox{NML}}}^{(N+1)}(y^t|x^{N+1})}{p_{\tiny{\mbox{NML}}}^{(N+1)}(y^{t-1}|x^{N+1})}
\end{align}
where the well known normalized maximum likelihood assignment for the entire sequence is 
\begin{align}
p_{\tiny{\mbox{NML}}}^{(N+1)}(y^N,y|x^N,x) = p_{\tiny{\mbox{NML}}}^{(N+1)}(y^{N+1}|x^{N+1}) = \frac{p_{\hat{\theta}(x^{N+1},y^{N+1})}(y^{N+1}|x^{N+1})} {\sum_{\tilde{y}^{N+1}}p_{\hat{\theta}(x^{N+1},\tilde{y}^{N+1})}(\tilde{y}^{N+1}|x^{N+1})}.
\end{align}

Before moving on, we note that in order to actually use this kind of probability assignment, the learner has to have some order for the data features sequence $x^N$. This is necessary since the data is not given with specifications regarding its order. Now, it is straightforward to show that any order function does not effect the results, yet the ordering is important for the case where there are some recurrent data features in $x^N$. Thus, the bound achieved by the scheme (\ref{NML}) only holds when all the data features in $x^N$ are different. 

Intuitively, this probability assignment should be sub-optimal, since it only takes into account at each time $t$ some of the observed outcomes that occur prior to $t$ in the chosen ordering. Nevertheless, its regret clearly upper bound the min-max optimal regret, i.e., we get the following upper bound over the min-max regret:

\begin{align}
\label{NML_bound}
\begin{split}
R^* &\leq R(q_{\mbox{\tiny{NML}}}(y_t|y^{t-1},x^{N+1}),z^{N+1}) =  
\frac{1}{N+1}\sum_{t=1}^{N+1} \log\left(\frac{p_{\theta(y^{N+1}|x^{N+1})}(y_t|x_t)}{q_{\mbox{\tiny{NML}}}(y_t|y^{t-1},x^{N+1})}\right) = \\
& =\frac{1}{N+1}\log\left(p_{\theta(y^{N+1},x^{N+1})}(y^{N+1}|x^{N+1})\right) - \frac{1}{N+1}\log\left(p_{\mbox{\tiny{NML}}}^{(N+1)}(y^{N+1}|x^{N+1}\right) = \\ &=\frac{1}{N+1}  \log\left(\sum_{\tilde{y}^{N+1}}p_{\theta(\tilde{y}^{N+1},x^{N+1})}
(\tilde{y}^{N+1}|x^{N+1})\right).
\end{split}
\end{align}

This result may be considered as the analogous to the $\frac{1}{N+1}I(Y^{N+1};\theta|x^{N+1})$ we got in the stochastic case, since this bound is essentially the loss incurred when trying to predict the whole sequence divided by the length of the sequence. Indeed, in the context of universal prediction, there are some connections between $I(Y^{N+1};\theta)$ and $\frac{1}{N+1}  \log\left(\sum_{\tilde{y}^{N+1}}p_{\theta(\tilde{y}^{N+1})}(\tilde{y}^{N+1})\right)$, as discussed in \cite{rissanen1996fisher}.

Another way to bound the leave-one-out minmax value is to use the pNML predictor for each time point $t$ while the rest $(N+1)\setminus t$ time points are the training. The resulting average leave-one-out regret for the entire sequence is given by:

\begin{align}
\label{pNML_bound}
R(q_{\mbox{\tiny{pNML}}},z^{N+1}) = \frac{1}{N+1}\sum_{t=1}^{N+1}\log\left(\sum_{\tilde{y}_t} p_{\theta(y^{{(N+1)} \setminus t},\tilde{y}_t,x^{N+1})}(\tilde{y}_t|x_t)\right) =
 \frac{1}{N+1}\log\left(\sum_{\tilde{y}^{N+1}}p_{\theta(y^{{(N+1)} \setminus t},\tilde{y}_t,x^{N+1})}(\tilde{y}_t|x_t)\right)
\end{align}

This expression can be interpreted as the empirical average regret of the pNML, $R^*_{ind}(z^{{N+1} \setminus t},x_t)$, over all possible choices of $t$. In addition, it is clear that maximizing this expression over $y^{N+1}$ yields an upper bound over the minmax regret of (\ref{loo_regret}).

Summarizing, both (\ref{NML_bound}) and (\ref{pNML_bound}) provide upper bounds on the leave-one-out regret which has a term that goes to zero as $O(1/N)$ for any sequence, provided that the model class is such that the numerator in both bounds is finite. In this case, the model class can be considered ``learnable''.



%% file: section_pnml_interpretation.tex
The pNML solution (\ref{individual_solution}) will be advocated as an alternative to the Empirical Risk Minimization (ERM) that is the most common approach in supervised learning. Its justification as the solution to the minmax problem (\ref{pnml_regret}) was discussed above. We propose in this section another interpretation, which will also lead to possible extensions of the proposed universal learner.

The learner we are looking for assigns a probability $q(\cdot|x;z^N)$ to the unknown test label $y$, given the test feature $x$ which can (and actually should) depend on the available training $z^N$. A possible way to describe the role of the training data is that it defines a subclass $P_{\Theta(z^N)}$ of models out of the larger model class $P_\Theta$ that fit, or comply with the training data. Once the model class is focused, the universal learner for the test data chosen to compete with any model in $P_{\Theta(z^N)}$. 

In particular, in the stochastic setting one may then solve
\begin{equation}
\label{minmax_sto_train}
\min_q \max_{p\in P_{\Theta(z^N)}} E \log \frac{p}{q} = \min_{q(y|x;z^N)} \max_{p(y|x) \in P_{\Theta(z^N)}} \sum_y p(y|x) \log \frac{p(y|x)}{q(y|x;z^N)}
\end{equation} 
where we also recognize $\sum_y p(y|x) \log \frac{p(y|x)}{q(y|x;z^N)}=D(p||q)$. Clearly, the min-max solution of (\ref{minmax_sto_train}) is a mixture within the refined class $\P_\Theta(z^N)$
\begin{equation}
q(y|x;z^N) = \int_{\theta\in \Theta(z^N)} w(\theta) p_\theta (y|x) d\theta
\end{equation}	
and $w(\theta)$ over $\Theta(z^N)$ maximizes $I\left(\Theta(z^n);Y|x\right)$.

In the more interesting individual setting the criterion will be
\begin{equation}
\min_q \max_y \max_{p\in P_{\Theta(z^N)}} \log \frac{p}{q} = \min_{q(y|x;z^N)} \max_y \max_{p(y|x) \in P_{\Theta(z^N)}} \log \frac{p(y|x)}{q(y|x;z^N)}
\end{equation} 
And if we now define 
\begin{align}
\label{gen_reference}
\hat{\theta}(z^N;y|x)=\arg\max_{\theta \in \Theta(z^N)} p_\theta(y|x),
\end{align}
then the criterion becomes
\begin{equation}
\label{minmax_ind_train}
\min_{q(y|x;z^N)} \max_y \log \frac{p_{\hat{\theta}(z^N,y|x)}(y|x)}{q(y|x;z^N)}
\end{equation}
By standard equalizer reasoning, the solution for $q$ in (\ref{minmax_ind_train}) is given by: 
\begin{align}
\label{generalized_pnml}
q(y|x; z^{N}) = \frac{p_{\hat{\theta}(z^N;y|x)}(y|x)}{\sum_{y'}p_{\hat{\theta}(z^N;y'|x)}(y'|x)}.
\end{align}

This solution is very similar to the pNML (\ref{individual_solution}). The difference is by how $\hat{\theta}$ is defined. In pNML the reference performance is given by (\ref{reference}), as the parameter value that fits (by maximizing the likelihood of) both the training and the test with the new possible label. In (\ref{generalized_pnml}), however, the reference $\hat{\theta}$ of (\ref{gen_reference}) fits only the test, but is chosen out of the subclass $P_{\Theta(z^N)}$ of parameter values deduced by the training. 

The question now is how to reasonably choose $P_{\Theta(z^N)}$. In \cite{PainskyFeder1}, that considered the stochastic setting, the class was chosen in the spirit of ``confidence intervals'', so that the true model will be in the class with, say, 95\% confidence. We suggest another natural criterion: the class will contain all models whose likelihood on the training is at least some value $c$, i.e.,

\begin{align}
\Theta(z^N) = \left\{\theta \in \Theta : p_\theta(y^N|x^N) \geq c  \right\}
\end{align}

By using a Lagrange multiplier $\lambda$ for this constraint, the choice in (\ref{gen_reference}) becomes
\begin{align}
\label{gen_reference_likelihood}
\hat{\theta}(z^N;y|x)=\arg\max_{\theta \in \Theta} \log p_\theta(y|x) + \lambda \log p_\theta(y^N|x^N)
\end{align}
Clearly, if the threshold $c$ is such that $\lambda=1$ then the resulting solution is the pNML.
Furthermore, the learner (\ref{generalized_pnml}) with the corresponding $\hat{\theta}(z^N;y|x)$ of (\ref{gen_reference_likelihood}) is a spectrum of learners where at $\lambda=\infty$ only the training is considered and the learner becomes the ERM, at $\lambda=1$ it is the pNML and at $\lambda=0$ the training is completely ignored and the resulting universal learner is useless. 

The approach and interpretation presented above seems to remind the setting suggested in \cite{TseFarnia}. In that work the training is also used to come up with a class of plausible models, and then the learner is solving a min-max problem. However, there are important differences. First, the class of models chosen by the training usually contain models for which some moments of the distribution comply with the empirical moments. Second, the quantity to min-max is the expected loss, not the regret, and third, since expectation is considered, the setting is stochastic and not individual. Nevertheless, the understanding that the role of the training is not to choose a single model, like the ERM, but to serve as a ``filter'' to unacceptable models appears in both works.

\subsection*{pALG} 
We now present another extension of the pNML, which has also a structure similar to (\ref{generalized_pnml}). Suppose there is a general procedure {\em ALG}, which can take a set $z^L=(x^L,y^L)$ and generate a value of a parameter $\theta\in \Theta$. It may be the maximum likelihood, but it may also be a regularized version of it with various penalty function, or any other procedure, like the result of an iterative algorithm, a stochastic gradient with some meta parameters that define the batch size and the number of iterations and so on. If we denote the outcome of the algorithm $\hat{\theta}_{\mbox{\tiny{ALG}}}(z^L)$, then the proposed pALG is given by:
\begin{align}
\label{palg}
q_{\mbox{\tiny{pALG}}}(y|x; z^{N}) = \frac{p_{\hat{\theta}_{\mbox{\tiny{ALG}}}(z^N,x,y)}(y|x)}{\sum_{y'}p_{\hat{\theta}_{\mbox{\tiny{ALG}}}(z^N,x,y')}(y'|x)}.
\end{align}

As for the pNML, the logarithm of the normalization factor, $\Gamma_{\mbox{\tiny{ALG}}}(z^N,x) = \log \sum_{y'}p_{\hat{\theta}_{\mbox{\tiny{ALG}}}(z^N,x,y')}(y'|x)$, which depends on the training $z^N$ and the test feature $x$, can be used to assess, locally, the learnability of the proposed method. Similarly to the discussion above regarding the learnability measure of the pNML, here, too, a small $\Gamma$ can attest on the stability and hence learnability of the proposed learning algorithm, while a large $\Gamma$ indicates uncertainty and large universality overhead. 



%% file: section_simulation_results.tex
\label{SimulationResults}
In this section we will present some simulation results for the universal solution to the individual learning problem (\ref{individual_solution}), and its respective regret (\ref{individual_regret}).

We will focus our attention to the simple hypotheses class of 1-d barrier threshold, given by:

\begin{align}
\label{1d_barrier_threshold}
p_{\theta = (b,p_1,p_2)}(y_t =1 |x_t)=
\begin{cases} p_1 &\mbox{ if } x_t \leq b \\
p_2 & \mbox{ if } x_t > b \end{cases}
\end{align}

In the following figures we present the average regret as function of $x_N$, where the average is taken over different realizations of the training set $z^{N-1}$. The data features in the training set are generated independently by a uniform distribution $x_t \sim U(0,1)$, and the outcomes are generated by some hypothesis in the hypotheses class $\theta = (b,p_1,p_2)$.

\begin{figure}[H]
	\begin{subfigure}{.5\textwidth}
		\centering
		\includegraphics[width=.9\linewidth]{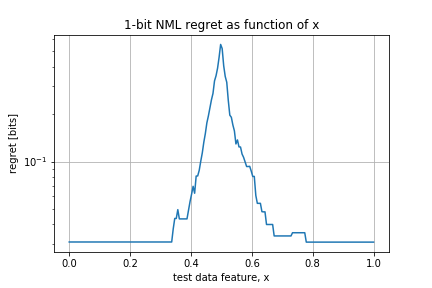}
		\caption{$\theta=(0.5,0.2,0.8), N = 100$, 100 runs}
		\label{fig_small_N}
	\end{subfigure}%
	\begin{subfigure}{.5\textwidth}
		\centering
		\includegraphics[width=.9\linewidth]{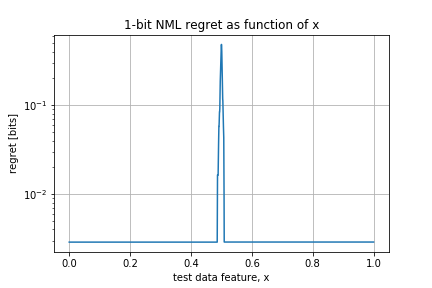}
		\caption{$\theta=(0.5,0.2,0.8), N = 1000$, 100 runs}
		\label{fig_large_N}
	\end{subfigure}
	\caption{Regrets as function of $x_N$ for different values of $N$}
	\label{fig_different_values_of_N}
\end{figure}

Figure (\ref{fig_small_N}),(\ref{fig_large_N}), show the impact of the number of samples on the regret.

Note that for both values of $N$ there is an area around $b$ for which the regret is relatively high. Those are areas in which the new label may change $b_{opt}$, thus changing the probability assignment for the next label significantly. Naturally, this uncertainty area decreases as $N$ increases, as there are fewer of values of $x_N$ for which the $b_{opt}$ changes. In addition, the regret outside those areas reduces as $N$ grows, since the effect of another label on the probability assignment decreases as $\frac{1}{N}$.

\begin{figure}[H]
	\begin{subfigure}{.5\textwidth}
		\centering
		\includegraphics[width=.9\linewidth]{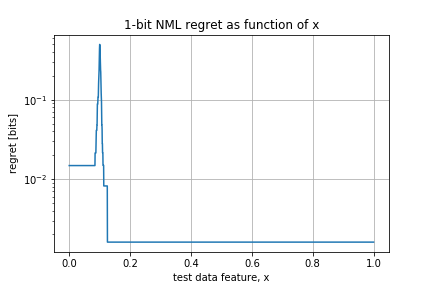}
		\caption{$\theta=(0.1,0.2,0.8), N = 1000$, 100 runs}
		\label{fig_b01}
	\end{subfigure}%
	\begin{subfigure}{.5\textwidth}
		\centering
		\includegraphics[width=.9\linewidth]{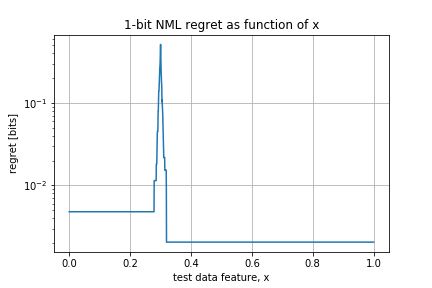}
		\caption{$\theta=(0.3,0.2,0.8), N = 1000$, 100 runs}
		\label{fig_b03}
	\end{subfigure}
	\caption{Regrets as function of $x_N$ for different values of $b$}
	\label{fig_different_values_of_b}
\end{figure}

One may note that as $b$ moves further from $0.5$, the regret in the area to the left of $b$ increases, which is due to the fact that the number of examples relevant to that area decreases.


%% file: conclusions.tex
\label{conclusions}
In this paper we have presented two possible definitions for the problem of supervised ``batch'' learning in the individual setting with respect to the logarithmic loss function. The first considers the typical situation where after observing $N$ training examples, $z^N$, there is a test sample, and we have come up with a criterion for finding the universal learner in that case. The second, considers a leave-one-out performance measure, justified by the fact that the reference learner is also of this type. We have presented the min-max solution to first problem, which we denoted as the predictive normalized maximum likelihood (pNML). In addition several possible upper bound for the second, leave-one-out problem, were also derived.

It is our thesis that the pNML learner is a valid alternative to the common ERM learner. It is more robust and it is strongly related to the stability feature that implies good generalization, in the proposed learner. In that respect, the pNML structure is more general and can be utilized for any learning algorithm, not just the one that maximizes the likelihood, as a scheme to stabilize that algorithm. We also note that our current research interests include the evaluation of the pNML for various learning problems and hypotheses classes such as linear regression and deep neural networks. These works are reported in \cite{pNML_linear_regression,pNML_neural_networks}.

In our opinion it will be interesting to find under what ``local'' conditions on the model class the pNML regret is small.  Also, it will be interesting to show the following conjecture: when the pNML regret is small, the empirical risk minimizer over the training set, which is the common way of choosing an hypothesis in real life applications, has a good generalization error. 

As for the leave-one-out formulation, clearly it is desired to solve the open problem of finding the corresponding min-max optimal solution.
And as for the pNML, to find out under what conditions on the model class (which may be ``local'') this regret is small, say going down to zero as $O(1/N)$ for any individual data. 

Finally, while mentioned briefly, in Appendix \ref{appendix b}, we considered a twice universal solution for the pNML. The twice universal approach is natural when the model class is unknown too, but there are several possible model classes, sometimes in a nested hierarchy. As appear in some of our current experiments, the twice universal approach show good performance in these unknown class cases.


%% file: AppendixA.tex
\label{appendix a}

In the stochastic setting, the performance of a
learner $q$ is evaluated by its expected {\em regret} $E \{ -\log q -
(-\log p_\theta)\} = E \log\frac{p_\theta}{q}$, where the expectation
is taken with respect to the true unknown probability. Written
explicitly and averaged over the training to accomodate for various
possible examples:
\begin{equation*} 
R_N(\theta,q;x^{N}) = \sum_{y^{N-1}\in {\cal Y}^{N-1}} p_\theta(y^{N-1}|x^{N-1}) R_N(\theta,q;x_N,z^{N-1}) = 
\sum_{y^N\in {\cal Y}^N} p_\theta(y^N|x^N)  \log\frac{p_\theta(y|x_N)}{q(y|x_N;z^{N-1})}
\end{equation*}
The learner $q$ depends on the training data, and its goal is to
minimize $R_N$. It does not know $\theta$, so it tries to make a good
choice, no matter what $\theta$ is. 
Thus, as proposed and analyzed in \cite{FogelFederISIT2018}, the learner $q$ is chosen by solving a min-max problem
\begin{equation}
\label{minmaxR} \min_q \max_{\theta} R_N(\theta,q; x^N) \end{equation}
One may further assume that features are also random and average $R_N$
over the possible $x^N$'s. 

The solution of (\ref{minmaxR}) is a learner which is a Bayes-mixture
over the hypotheses class,  
\begin{align}
\label{Bayesian_solution}
q(y|x_N;z^{N-1}) = \int_{\theta} \frac{w(\theta)p_\theta(y^{N-1}|x^{N-1})}{\int_{\theta'}w(\theta')p_\theta(y^{N-1}|x^{N-1}) d\theta'}\; p_\theta(y|x_N) d\theta 
= \int_{\theta} w(\theta | z^{N-1}) p_\theta(y|x_N) d\theta
\end{align}
where as in \cite{davisson1980source}, following a
``redundancy-capacity'' theorem, $w(\theta)$ maximizes
$I(\Theta;Y_N|Y^{N-1},x^N)$, the mutual information between the class
and $Y_N$ given the previous labels $Y^{N-1}$ (all are random
variables) and conditioned on $x^N$.  
Intuitively, models that have high a-posteriori probability given the
training are weighted accordingly in the final learning outcome. The
resulting maximal mutual information, or ``capacity'', may be
considered as a measure of the learnability of the model class.


%% file: AppendixB.tex
\label{appendix b}

Let us consider a setting where one is given a set of hypotheses classes, $\Theta_1,...,\Theta_k$, and is required to attain good performances with respect to all of them. Following previous results from universal prediction, we will consider the following batch solution:

\begin{equation} 
	q(y|x; z^{N}) = \frac{\max_i q_i(y|x;z^N)}{\sum_{y'\in {\cal Y}}\max_{i} q_i(y'|x,z^N)}
\end{equation}

where

\begin{equation}
\label{universal_i}
	q_i(y|x; z^N) = \frac{p_{\hat{\theta_i}(z^N,x,y)}(y|x)}{\sum_{y'\in {\cal Y}}p_{\hat{\theta_i}(z^N,x,y')}(y'|x)}.
\end{equation}

Denoting $j = j(y) = \argmax_i p_{\hat{\theta_i}(z^N,x,y)}(y|x)$ we get the following bound over the regret:

\begin{align}
\label{regret_TW1}
\begin{split}
R = R(z^N,x,y) &= \log{\frac{p_{\hat{\theta_j}(z^N,x,y)}(y|x)}{q(y|x; z^N)}} = \\
&=
 \log{\frac{p_{\hat{\theta_j}(z^N,x,y)}(y|x)}{\max_i q_i(y|x; z^N)}} + \log{\sum_{y'\in {\cal Y}}\max_{i} q_i(y'|x;z^N)} \leq \\
 & \leq 
 \log{\frac{p_{\hat{\theta_j}}(z^N,x,y)(y|x)}{ q_j(y|x; z^N)}} + \log{\sum_{y'\in {\cal Y}}\max_{i} q_i(y'|x;z^N)} = \\ 
 &= \log{\sum_{y'\in {\cal Y}}p_{\hat{\theta_j}(z^N,x,y')}(y'|x)} + \log{\sum_{y'\in {\cal Y}}\max_{i} q_i(y'|x;z^N)} = 
 \overline{R}(z^N,x)
\end{split}
\end{align}
where while the regret $R$ may also depend on the value of the true label $y$, we have bounded it by the quantity $\overline{R}$ which does not depend of that $y$. 

Interestingly, $\overline{R}$ has two terms, where the first term represents the log-normalization factor, or the ``learnability'' with respect to the class $\Theta_i$, which correspond to the best fit of the model and class to the true label. The second term is the cost of the twice universal procedure and it is at most $\log k$. Unfortunately, sometimes, $\log k$ is a too large cost to pay. But, since this term can be easily evaluated explicitly for the specific training and test, it may be small in many cases; then the twice universality gain is attained with no much cost.

There is yet another way to assess the twice universality solution. Denote now by $k=k(y)=\arg \max_i q_i(y|x;z^N)$ where $q_i$ is the universal probability within $\Theta_i$ given by (\ref{universal_i}).
We now define the twice universal regret as
\begin{equation} 
\tilde{R} = \log{\frac{q_k(y|x;z^N)}{q(y|x; z^N)}} = \log{\sum_{y'\in {\cal Y}} q_{k(y')} (y'|x;z^N)}
\end{equation}
This term is similar to the second term in the regret (\ref{regret_TW1}), by using $k(y)$ instead of $j(y)$. Since the regret is with respect to $\max_i q_i$, we note that $q_i$ contains already the overhead of the class $\Theta_i$.
